\documentclass{optica-article}

\journal{opticajournal}

\articletype{Research Article}
\usepackage{xcolor}
\usepackage{lineno}
\usepackage{float}


\AtBeginDocument{\captionsetup{width=\textwidth}}

\begin{document}

\title{OAM Light Demultiplexing from an Intensity Profile using Orthogonality Restoration of Pair Modes}

\author{Junsu Kim\authormark{1}, Hyunchae Chun\authormark{2*}, and SeungRyong Park\authormark{1*}}

\address{\authormark{1,1*}Department of Physics, Incheon National University (INU), Incheon, South Korea}

\address{\authormark{2*}Department of Information and Telecommunication Engineering, Incheon National University (INU), Incheon, South Korea}

\email{\authormark{1}esterbk@inu.ac.kr, \authormark{2*}hyunchae.chun@inu.ac.kr, \authormark{1*}abepark@inu.ac.kr}

\begin{abstract*}
Orbital Angular Momentum (OAM) of light is a promising degree of freedom for next-generation communication. By exploiting the orthogonality of OAM modes, multi-channel division enables a linear increase in communication performance proportional to the number of available modes. However, the multiplexing and demultiplexing of each superposition state remain essential yet complex processes. Demultiplexing has been established through spatial-domain methods that require additional optical elements such as gratings and apertures, which can decrease communication efficiency and accuracy under various conditions. In this paper, we propose a demultiplexing method under a single intensity profile by orthogonality restoration of OAM pair states. This method can be applied directly to an OAM multichannel communication system without additional receiver-side optical structure. We present simulation results of our method under various conditions.
\end{abstract*}

\section{Introduction}
Light can carry Orbital Angular Momentum (OAM) \cite{Allen1992,Poynting1909,Yao2011,Padgett2017}, a physical quantity that finds its mathematical representation in the Laguerre-Gaussian (LG) mode expansion. Such LG beams exhibit a characteristic vortex-like wavefront pattern, commonly referred to as optical vortices or twisted photons \cite{Stoyanov2023,MolinaTerriza2007,Neu1990a,Neu1990b,Baumann2009,Vickers2008,Paufler2019,Erhard2017,Mair2001}. The theoretical foundation for these modes lies in the paraxial approximation of the Helmholtz equation, solved in cylindrical coordinates with Laguerre polynomials $L^{|\ell|}_n$ that provide the radial dependence. The detailed form of the LG mode wavefunction is given by
\begin{equation}
\begin{aligned}
    \psi_{n,\ell}(r,\phi,z)\propto\frac{w_0}{\sqrt{2}r}W^{|\ell|+1} L^{|\ell|}_n(W^2)\ \exp\left[-\frac{W^2}{2}+i\frac{z}{2z_0}W^2-i(2n+|\ell|+1)\tan ^{-1}{\frac{z}{z_0}}\right]\ \exp{(-i\ell\phi)}
\end{aligned}
\end{equation}
where $W\equiv W(r,z)=\frac{\sqrt{2}r}{w(z)}$, $z_0$ is the Rayleigh range, $\ell$ is the topological charge, $n$ is the radial mode index, $\phi$ is the azimuthal angle, and $w(z)=w_0\sqrt{1+(z/z_0)^2}$ with the beam waist $w_0$. Its physical interpretation is well established. A notable feature of this structure is the azimuthal phase dependence, manifesting as the complex exponential term $e^{-i\ell\phi}$. The integer $\ell$ corresponds to the topological charge, which quantifies the OAM content and directly influences how the phase varies around the beam axis \cite{Leach2002,Krenn2018,Kotlyar2017}. Optical communication technologies leverage the orthogonality between different OAM states, enabling parallel information channels \cite{Willner2015,Willner2021,Trichili2019,Garcia2008}. Achieving the full potential of such systems requires robust methods for combining and separating different OAM channels, in processes known as multiplexing and demultiplexing (mux and demux). These operations are critical for maintaining signal integrity and maximizing data throughput in modern optical communication systems \cite{Wang2012,Huang2014,Bozinovic2013,Gibson2004,Yan2014,Gong2019,Ni2025,Shi2022}.

The orthogonality of each OAM state is a key feature that enables mux-demux operations. This property is mathematically defined by the inner product between distinct states as
\begin{equation}
\langle \psi(\phi),\psi^{\prime}(\phi)\rangle=\int \psi^*(\phi)\ \psi^{\prime}(\phi)\ d\phi=0
\end{equation}
The electric field of an ideal OAM mode can be expressed simply, and orthogonality between different states ($\ell\neq \ell^{\prime}$) is demonstrated as
\begin{equation}
    \mathbf E_{\text{OAM}}\propto e^{-i\ell\phi}\to\int_0^{2\pi} e^{+i\ell\phi}\ e^{-i\ell^{\prime}\phi}d\phi=2\pi\delta_{\ell-\ell',0}
\end{equation}
However, this property holds only in the electric field domain. The intensity of light is represented as $|\mathbf E|^2$, and the modes lose orthogonality with each other: $\int|\mathbf E_{\ \ell}|^2|\mathbf E_{\ \ell^{\prime}}|^2 d\phi\neq0$. The orthogonality of each multiplexed OAM state is not conserved in a single intensity profile.

Consequently, mux-demux processes are generally performed in the spatial domain using additional elements such as diffraction gratings \cite{Lei2015,Zheng2017,Leith1962,Heckenberg1992,Stoyanov2015,Janicijevic2008,Topuzoski2011,Topuzoski2019,Topuzoski2009,Karahroudi2017}, meta-surfaces \cite{Ren2019,Yu2024a,Yu2024b}, refractive modulators \cite{Beijersbergen1994,Longman2017,Arlt1998,Junsu2025}, and mode sorters \cite{Berkhout2010,Lavery2012,Mirhosseini2013,Fontaine2019} before detecting the intensity of light. Computational approaches using neural networks have also been explored \cite{Ye2024,Wang2021,Geng2024,Fang2022}. These methods introduce additional steps and device layout constraints for demux process. Atmospheric turbulence poses an additional challenge to OAM based communication by inducing inter-modal crosstalk \cite{Paterson2005,Ren2013,Malik2012,Manavalan2025,Chen2025,Liu2026}, and its compensation remains an active area of research. To address the structural complexity inherent in these spatial domain approaches, we propose an intuitive demux method under a single intensity profile without additional optical processes. To recover orthogonality among the OAM modes, we define orthogonality restoration as the subtraction of an offset at each radius from the intensity profile, which recovers both positive and negative phase values. In our scheme, an OAM pair mode becomes a single channel. This approach introduces an intensity channel based on OAM pair modes, which is fundamentally orthogonal to each other, thereby enhancing the aggregate data-rate by a factor proportional to the number of available OAM modes without requiring additional bandwidth or temporal resources. The details of the pair mode theory are presented in the next section, and the radial distribution property of OAM is addressed in the simulation section.

The present work formulates intensity domain demultiplexing as an analytical orthogonality problem: by applying orthogonality restoration to the superposed pair mode profile, a single Fourier Transform recovers the continuous amplitudes $A_p$ of all multiplexed channels in parallel, yielding a multi-channel amplitude-modulated communication architecture with aggregate data-rate scaling linearly with channel count. This stands in contrast to recent works that employ the same $\cos^2(\ell\phi)$ azimuthal profile of $LG_{+\ell}+LG_{-\ell}$ superpositions but reduce reception to discrete pattern recognition or template correlation tasks. Examples include cross-correlation against a precompiled library of reference intensity templates~\cite{Rumman2026}, deep learning based recognition of flower-like intensity patterns~\cite{Chuo2025}, or hologram-correlation decoding for petal-OAM multiplexed encryption~\cite{Lin2025}. Whereas these template or learning based schemes operate in the regime of single mode recognition or image reconstruction, the analytical restoration framework introduced here removes the need for additional optical processes, reference templates, or trained models and addresses the simultaneous separation of multiple amplitude modulated channels from a single intensity frame, establishing a distinct paradigm of multi channel intensity domain communication.

\section{Theory}
\begin{figure}[H]
    \centering
    \includegraphics[width=\linewidth]{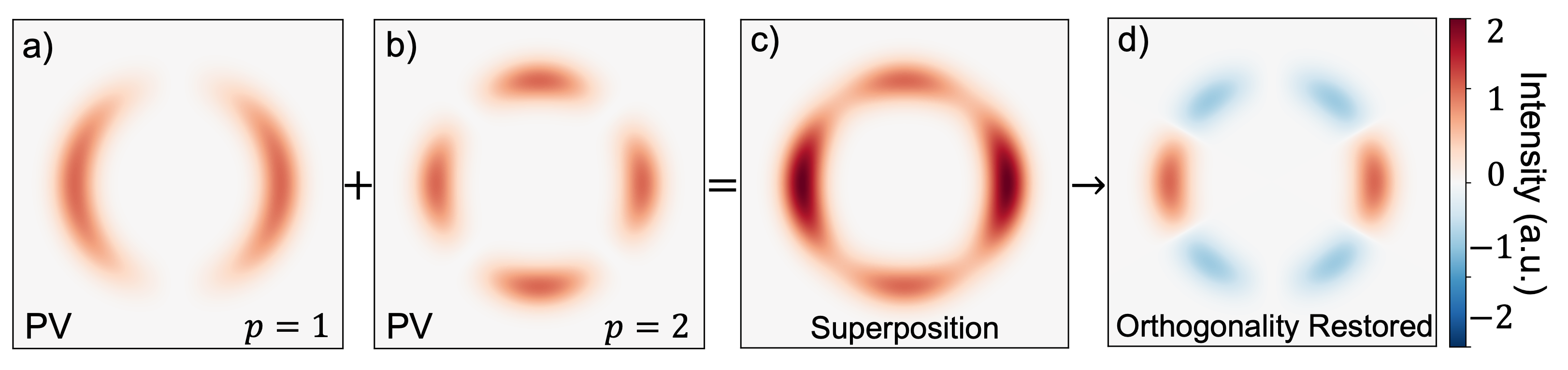}
    \caption{Example of Perfect Vortex (PV) OAM pair state intensity profile: a) $p=1$ state with $\ell=\pm1$ in superposition, b) $p=2$ state with $\ell=\pm2$ in superposition with 0 to 1 intensity scale, c) multiplexed state of pair $1,2$, d) restored profile obtained by subtracting half of the maximum at each radial position with -2 to 2 scale. The scale bar covers all figures. Through our proposed method, the detected superposition profile in c) is transformed into the restored profile in d).
    }
    \label{fig:pair_state}
\end{figure}

In this communication scenario, we consider the fundamental radial mode ($n=0$). The individual channel state is based on the OAM pair state $|p\rangle$, which is a superposition of topological charges with equal magnitude but opposite sign. It is given by
\begin{equation}
    |p\rangle\equiv \frac{1}{\sqrt 2}\left(|\ell\rangle+|-\ell\rangle \right)
\end{equation}
where $p=|\ell|$, with both $\ell$ and $p$ being non-zero integers. We designate this $|p\rangle$ as a communication channel. The electric field of this pair state is represented using simple trigonometric functions:
\begin{equation}
    \mathbf E_{p}\propto e^{-i\ell\phi}+e^{+i\ell\phi}=2\cos(\ell\phi)=2\cos(p\phi)
\end{equation}
where $\phi$ is the azimuthal angle. In the intensity profile, the pair state can be expressed as
\begin{equation}
    |{\mathbf E_p}|^2=I_p\propto\cos^2(p\phi)
\end{equation}
and its function plots are shown in Fig.~1. These $|p=1\rangle$ and $|p=2\rangle$ are not orthogonal to each other: $\int|\mathbf E_1|^2|\mathbf E_2|^2d\phi\neq0$. We propose an orthogonality restoration method that centers the oscillating intensity function on zero. The restored intensity profile is obtained by subtracting half of the maximum intensity from the original profile, at each given radius:
\begin{equation}
    I_p^{(\text R)}\equiv I^{\ \text{max}}_p\left[\cos^2(p\phi)-\frac{1}{2}\right]=\frac{I^{\ \text{max}}_p}{2}\cos(2p\phi)
\end{equation}
The resulting intensity profile contains negative values, as shown in Fig.~1(d). This restored pair (R-pair) intensity $I^{(\text{R})}_p$ satisfies the orthogonality property in the detection plane:
\begin{equation}
    \int I^{(\text R)}_p\ I^{(\text R)}_{p^{\prime}}d\phi\to\int_0^{2\pi}\cos(2p\phi)\cos(2p'\phi) d\phi=0
\end{equation}
where $p\neq p^{\prime}$. Therefore, we can demux from a single intensity profile completely with this orthogonality restoration method, as shown in Eq.~(8). The multiplexed OAM R-pair intensity profile can be represented as
\begin{equation}
    I_{\text{total}}^{(\text{R})}=\sum_{p}\frac{I^{\ \text{max}}_p}{2}\cos(2p\phi)
\end{equation}
We treat each pair channel as mutually incoherent, so that intensities sum directly without inter-channel interaction. Separating the orthogonal oscillation functions of each R-pair mode is carried out by a Fourier Transform (FT) process. With this, the weight of each multiplexed OAM mode is also detectable as a signal. This method can be applied directly to an OAM multichannel communication system without additional optical structure. However, in a practical detection scenario, we cannot directly detect the amplitude of each mode in a multiplexed pair state. We therefore proceed as follows: the intensity profile of phase-matched multiplexed pair states can be represented as
\begin{equation}
I_{\text{signal}}\equiv\sum_pA_p\cos^2(p\phi)=\sum_p\left[\frac{A_p}2+\frac{A_p}2\cos(2p\phi)\right]
\end{equation}
where $A_p$ is the amplitude of each pair state. This can be interpreted as an amplitude shift keying (ASK) signal communication scenario \cite{Proakis2008,Goodman2017}. The maximal intensity is obtained as $I^{\text{max}}_{\text{signal}}\equiv\sum_{p}A_p$ when $\phi=0$. For the $I^{\text{max}}_{\text{signal}}\neq\sum_{p}A_p$ case like phase error, the subtracted quantity is calculated as the mean value as $\frac{1}{2\pi}\int_0^{2\pi} I_{\text{signal}} d\phi=\frac12\sum_pA_p$. The maximal intensity is directly detectable, and we can rewrite Eq.~(10) as
\begin{equation}
 I_{\text{signal}}-\frac{I^{\text{max}}_{\text{signal}}}{2}=\sum_{p}\frac{A_p}2\cos(2p\phi)\equiv I^{(\text{R})}_{\text{signal}}
\end{equation}
by subtracting half of the maximum at each radius from the total intensity. The left-hand side of Eq.~(11) consists of detectable values. The subtracted quantity must be estimated locally at each radius, scaling with the radial intensity profile rather than as a single global constant. In the $I^{(\text{R})}_{\text{signal}}$ profile, orthogonality among the modes is restored. We extract the pair state $p$ and amplitude $A_p$ by FT, thereby obtaining complete information about the pair mode states.

This FT based mode recovery is inherently robust to phase errors between $\pm\ell$ states which manifest as a phase offset of each pair state. The phase error between $\pm\ell$ states can be defined as $\Delta\phi_p$ and the intensity profile of the multiplexed pair state is given by 
\begin{equation}
I_{\text{signal}}^{(\Delta\phi_p)}=\sum_p A_p\cos^2(p\phi+\Delta\phi_p/2)=\sum_p\frac{A_p}{2}[1+\cos(2p\phi+\Delta\phi_p)]
\end{equation}
We can see that the phase error between the $\pm\ell$ states is equivalent to a phase offset of each pair state. They do not affect the frequency and orthogonality of the restored profile, and the FT magnitude is independent of $\Delta\phi_p$. We can subtract the mean value at each radius of the intensity profile to obtain the restored profile as
\begin{equation}
\frac1{2\pi}\int_0^{2\pi}I_{\text{signal}}^{(\Delta\phi_p)}\ d\phi=\sum_p\frac{A_p}{2}\to I^{(\text{R},\Delta\phi_p)}_{\text{signal}}=\sum_p\frac{A_p}{2}\cos(2p\phi+\Delta\phi_p)
\end{equation}

The practical requirement is therefore not strict phase calibration but rather temporal phase stability within a single frame acquisition. 
There are limitations on the minimum detectable unit determined by resolution and signal purity. The detection plane has finite pixel size and bit depth in imaging sensors such as CCD and CMOS. The signal can be affected by noise and other perturbations. We present simulation results regarding the feasibility and limitations under white noise conditions to verify the validity of our method. We also consider a scenario that includes the radial distribution of the OAM beam, which is treated separately in the following subsection.

\section{Results and Discussion}
We design simulations for the R-pair demultiplexing process in the intensity profile with FT under communication conditions and estimate the available limiting conditions. A typical communication system consists of multiple channels, each carrying different information. In our simulation, we consider a scenario with 128 multiplexed OAM pair channels, where each channel carries a digital bit stream encoded via amplitude modulation.
\begin{figure}[!tb]
    \centering
    \includegraphics[width=\linewidth]{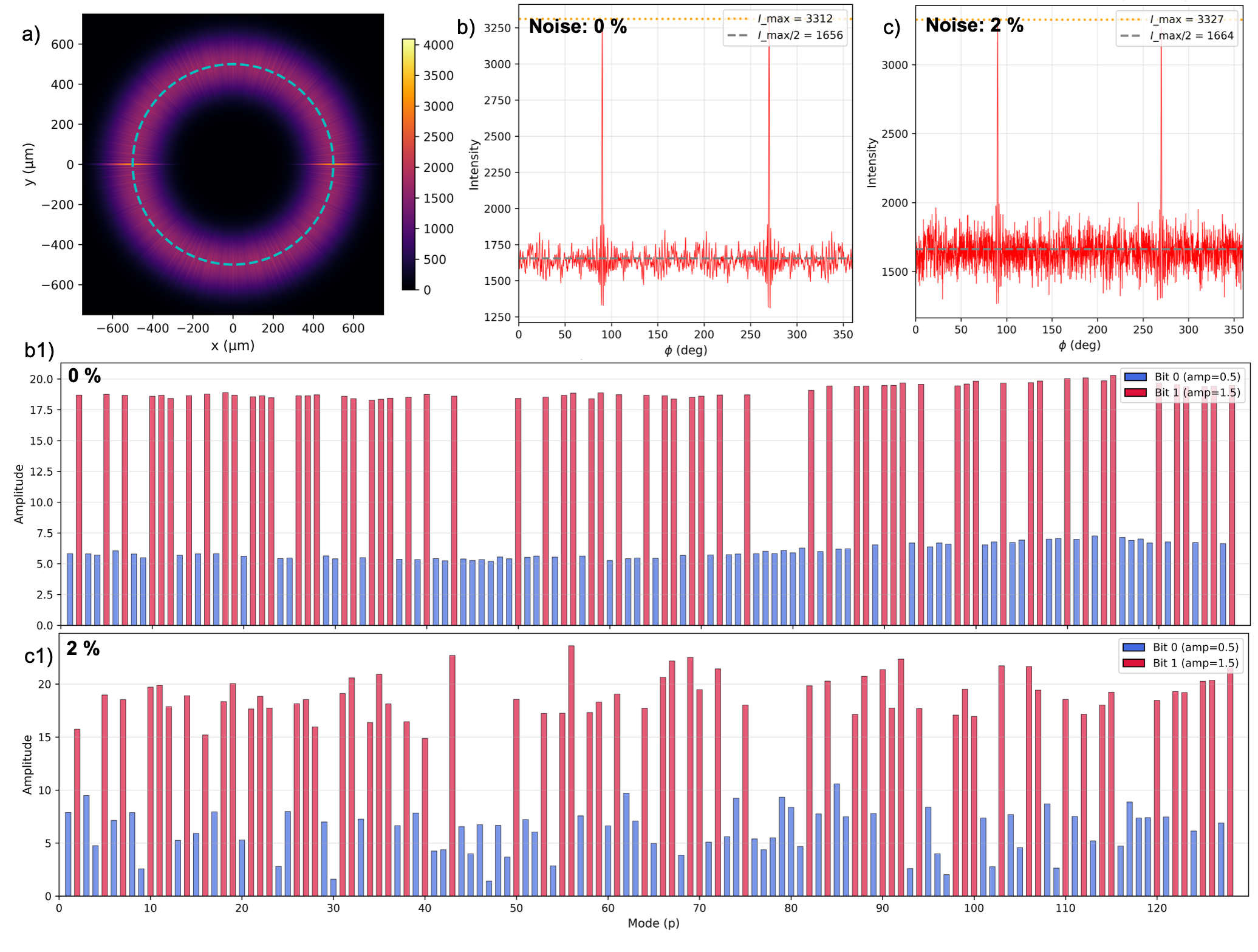}
    \caption{OAM pair mode 2-PAM (pulse amplitude modulation) ($\pm 0.5$) communication scenario with 128 multiplexed channels. a) CCD detected image of the multiplexed pair modes intensity profile ($1\ \mu$m pixel, 12-bit depth); the blue dashed line indicates the calculating radius ($r=500\ \mu$m). b),~c) Circular scan profile at the detection radius for the ideal 0\% and 2\% white noise cases, respectively. b1),~c1) Fourier Transform result of the detected signal for each condition.}
    \label{fig:pv_results}
\end{figure}

The number of pair states determines the aggregate data-rate of communication. Because all $N$ number of pair channels are transmitted and detected simultaneously within a single frame, the system operates as an inherently parallel communication link. The aggregate data-rate scales linearly with the number of pair mode channels for $M$-PAM (pulse amplitude modulation). For example, 128 pair channels at 2-PAM deliver 128 bits/frame, increase aggregate data-rate compared to a conventional single-mode link. We construct 2-PAM, 4-PAM, and 8-PAM communication systems by adjusting the amplitude of each pair state. In a 2-PAM system, each pair state has either 0 or 1 signal relative to amplitude. In a 4-PAM system, each pair state has either 0, 1/3, 2/3, or 1 signal relative to amplitude. In an 8-PAM system, each pair state has either 0, 1/7, 2/7, ..., or 1 signal relative to amplitude.
To verify the method, we construct a 2-PAM (1 bit per pair mode) system with $\pm 0.5$ amplitude range to differentiate symbols at the reference signal level. The simulation is based on a detection imaging scheme with 1~$\mu$m pixel size and 12-bit depth resolution CCD detector. Numerous system parameters affect signal validity, including pixel size, bit depth, noise, modulation order, and radial dependence. First, we present a perfect vortex (PV) scenario without OAM radial dependence of the intensity profile. Subsequently, we demonstrate the radial distribution property in a later subsection. For each scenario, white noise is induced to verify robustness, and we provide calculation results under various conditions in the last part of this section.

\subsection{Perfect Vortex Scenario}
The ideal condition of OAM communication is when all pair states are perfectly orthogonal and can be distinguished without any interference. OAM beams are constructed as Perfect Vortex (PV) \cite{Vaity2015,Liu2017,Pereiro2023} without radial dependence of intensity, which confirms the spatial minimum occupancy advantage of communication. In this scenario, the intensity profile of the multiplexed OAM light can be accurately reconstructed from the signal, and the FT process effectively separates the orthogonal oscillation functions of each R-pair mode. The communication data (TX, RX) is ``Junsu Kim Stream'' in bit stream format.

Fig.~2 shows the simulation results. The CCD detected image of the multiplexed pair modes intensity profile is shown in Fig.~2(a). The blue dashed line represents the radius at which the intensity profile is analyzed. The circular scan profiles at $r=500\ \mu$m are shown in Fig.~2(b) for the ideal case and Fig.~2(c) for the 2\% noise case. The FT results are presented in Fig.~2(b1) and Fig.~2(c1), where amplitudes correspond to the different OAM pair modes as bit signals.
The noise case (Fig.~2(c1)) shows decreased SNR (signal to noise ratio) compared to the ideal case, as expected. Nevertheless, the method achieves error free decoding (BER~$=0$) even under 2\% noise at 128 modes, demonstrating the robustness of the PV circular scan approach combined with the $I_{\max}/2$ subtraction and FT processing. The next subsection addresses the OAM radial distribution scenario, which extends the analysis using the same principle.

\subsection{Considering OAM Radial Distribution}
We can separate modes with OAM radial dependence by using the Radial Integration Method \cite{DErrico2017}. Without the PV approach, the circular scan method requires sampling at a different radius for each mode. An alternative approach is to integrate the intensity over the entire radial range, extracting state information in a single operation as
\begin{equation}
I_{\text{integrated}}(\phi) \equiv \int_0^{R} I(r,\phi) \cdot r \, dr
\end{equation}
where $r$ and $\phi$ are the radial and azimuthal coordinates in polar coordinates, and $R$ is the integration boundary. The intensity radial distribution of an LG mode beam profile with topological charge $\ell$ is given by
\begin{equation}
|u_\ell(r)|^2 = \left( \frac{\sqrt 2r}{w_{\ell}} \right)^{2|\ell|} \exp\left( -\frac{2r^2}{w_\ell^2} \right)
\end{equation}
where the local maximum is found at $d|u_\ell(r)|^2/dr=0\to r_{\text{max}} =w_\ell \sqrt{|\ell|/2}$. This is the radius of maximum intensity for mode $\ell$, and $w_\ell$ is the effective beam width parameter for mode $\ell$. Substituting the mux pair state intensity distribution $I(r,\phi) = \sum_p A_p |u_p(r)|^2 \cos^2(p\phi)$, we obtain
\begin{equation}
I_{\text{int}}(\phi) = \sum_p A_p \bar W_p \cos^2(p\phi)
\end{equation}
, where we define the weight of each mode as
\begin{equation}
\bar W_p \equiv \int_0^{R} |u_p(r)|^2 \cdot r \, dr
\end{equation}
Finally, we obtain
\begin{equation}
\begin{aligned}
I_{\text{int}}(\phi)=\sum_p \frac{A_p\bar W_p}{2}[1+\cos(2p\phi)]=I_{\text{mean}}+\sum_p \frac{A_p\bar W_p}{2}\cos(2p\phi)
\end{aligned}
\end{equation}
, where $I_{\text{mean}}\equiv \sum_p A_p \bar W_p/2$. The FT of the mean subtracted angular profile yields peaks at frequencies $2p$ with magnitudes $M(2p)\equiv A_p \bar W_p/2$. As each mode contributes proportionally to its weight $\bar W_p$, direct comparison of FT peak magnitudes also makes higher-order modes clearly detectable. To recover the true amplitude $A_p$, we calculate it using the pre-computed weight as
\begin{equation}
\tilde{A}_p = \frac{2|M(2p)|}{\bar W_p}
\end{equation}
where $\tilde{A}_p$ denotes the estimated amplitude from the measurement, ideally equal to the transmitted amplitude $A_p$. This process ensures that all modes are measured on an equal basis regardless of their spatial profiles. The method offers two practical advantages: noise is averaged over the full integration area, and all mode amplitudes are extracted simultaneously from a single FT operation. Integration over the radial profile effectively increases the measurement area, leading to improved SNR due to the larger number of pixels utilized compared to the PV single radius detection scenario. This approach may also introduce cost related drawbacks due to the increased detection area required. Nevertheless, considering the radial distribution can improve the communication capacity despite the complexity of requiring a large and high resolution sensor.

\begin{figure}[!tb]
    \centering
    \includegraphics[width=\linewidth]{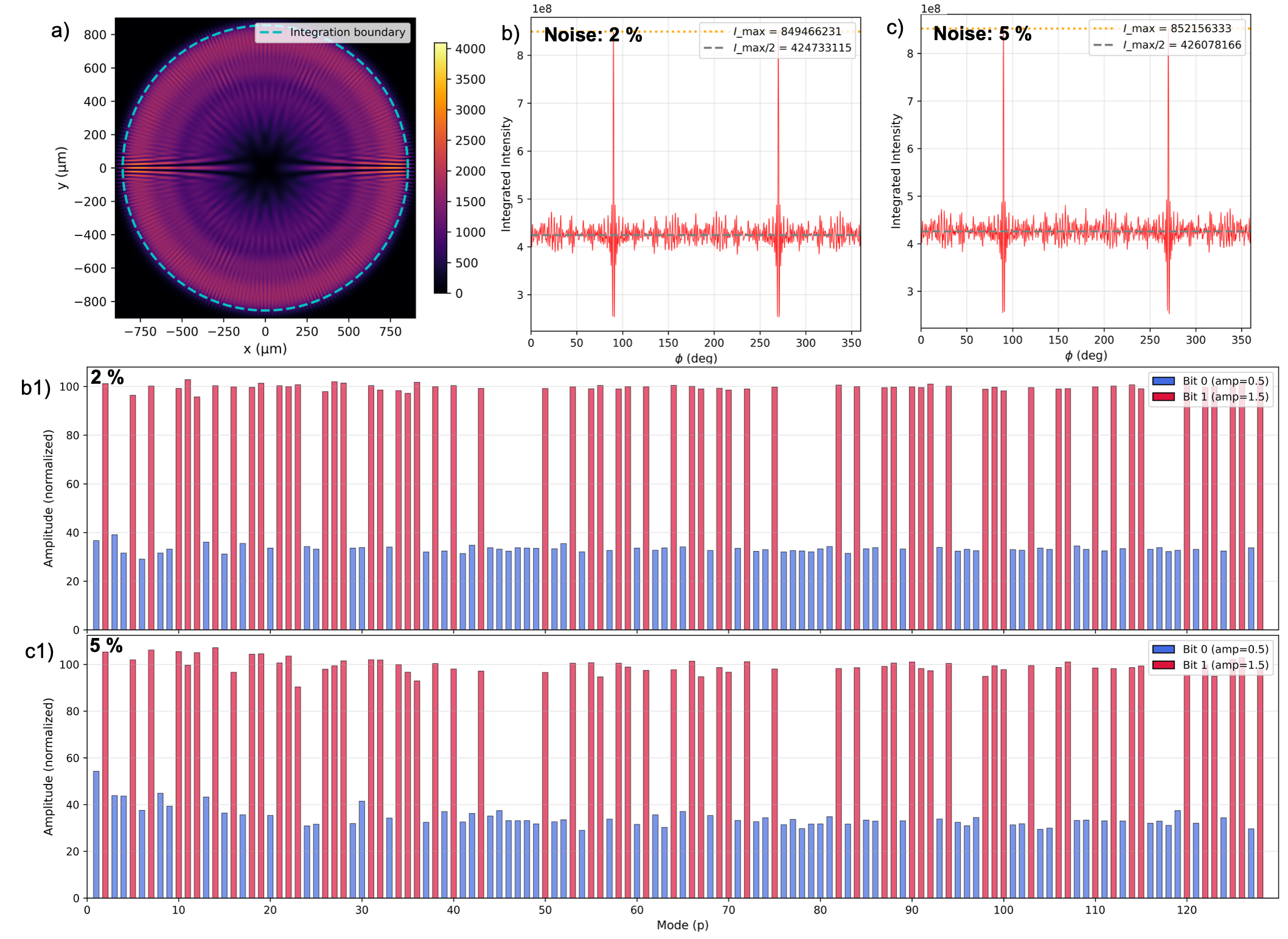}
    \caption{OAM pair mode 2-PAM ($\pm0.5$) communication scenario considering radial distribution with 128 multiplexed channels. a) CCD detected image of the multiplexed pair modes intensity profile ($1\ \mu$m pixel, 12-bit depth); the blue dashed line represents the integration boundary. b), c) Radial integration profiles for 2\% and 5\% white noise cases. b1),~c1), Fourier Transform results for 2\%, and 5\% white noise cases, respectively.}
    \label{fig:radial_results}
\end{figure}

Fig.~3 shows the scenario considering radial distribution under the same communication conditions. White noise 2\% case (Fig.~3(b1)) achieves BER~$=0$ and high SNR, demonstrating the advantage of integrating over the radial profile. The decoded bits remain distinguishable at both 2\% and 5\% noise levels, with quantitative BER performance detailed in Fig.~4. This robustness arises because the radial integration averages spatially uncorrelated noise over a large number of pixels, and the FT process concentrates the signal energy at discrete frequencies while distributing noise across all frequency.

\subsection{Validity of Our Method}
To quantify the performance boundaries, we calculate BER across a range of system parameters for both the PV and radial distribution considered method. Each BER value is averaged over 20 independent trials with pre-generated random bit streams shared across all parameter conditions, ensuring that observed differences arise solely from the varied parameter and not from statistical fluctuation of the transmitted data.

\begin{figure}[!tb]
    \centering
    \includegraphics[width=\linewidth]{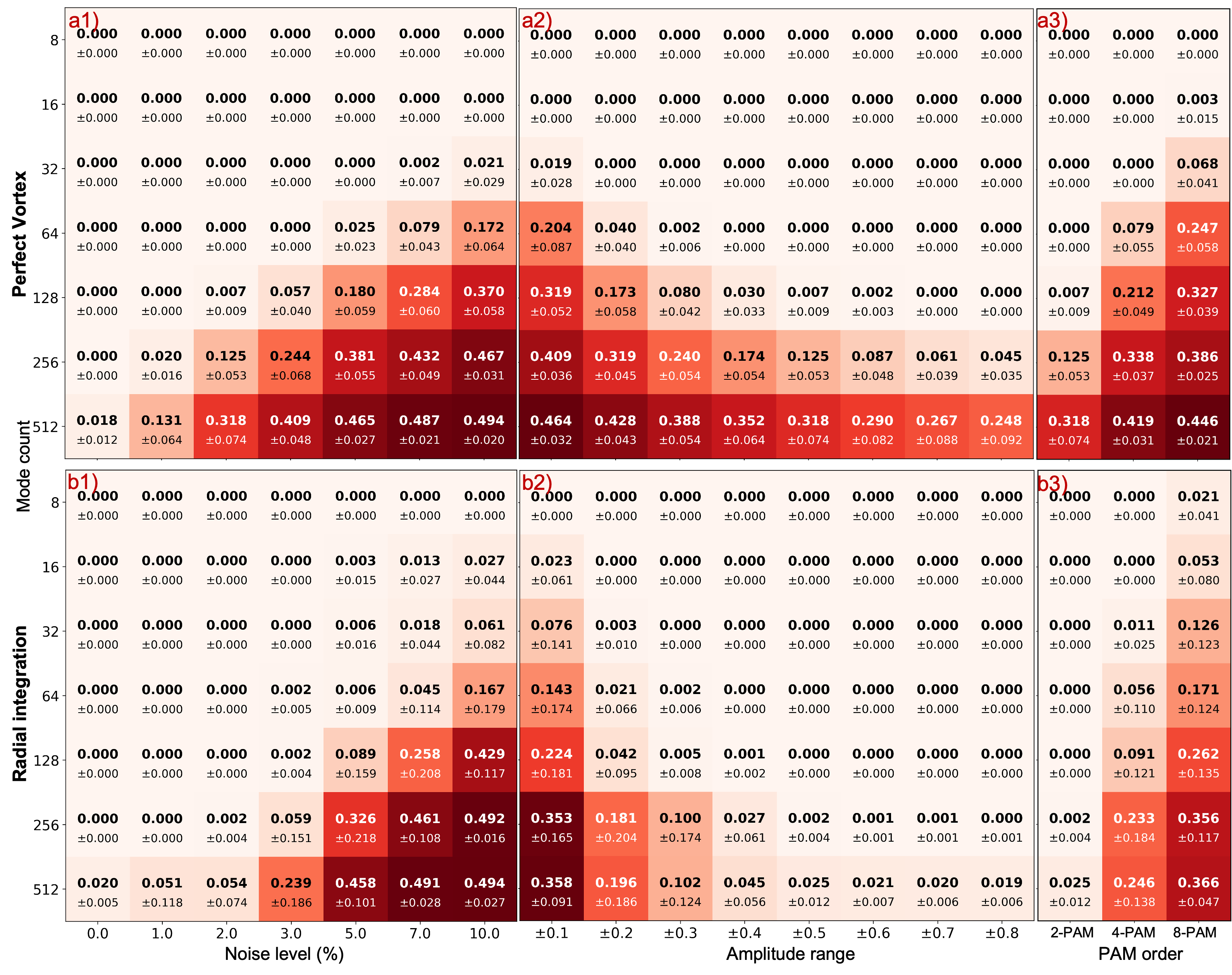}
    \caption{Bit Error Rate (BER) analysis for the PV scenario (a1--a3) and radial distribution considered method (b1--b3). a1,~b1) BER versus noise level (0--10\%) and pair mode count ($2^3$--$2^9$) at fixed 2-PAM amplitude range $\pm0.5$. a2,~b2) BER versus amplitude symbol range ($\pm0.1$--$\pm0.8$) and pair mode count at 2\% noise with 2-PAM. a3,~b3) BER versus modulation order (2/4/8-PAM) and pair mode count at 2\% noise with amplitude range $\pm0.5$. Each BER result is averaged over 20 repeated calculations with identical pre-generated random bit streams shared across all conditions. Standard deviations of 20 trials are also calculated.}
    \label{fig:ber_calc}
\end{figure}

Fig.~4(a1,~b1) shows that BER increases with both noise level and pair mode count, and also shows method's consistency in both PV and radial distribution considered scenarios. The variation of standard deviation for each conditions comes from fixed parameters of image sensor. And the increasing measurement area in radial integration, leads to low BER than PV in some extreme conditions. For instance, at 128 modes and 2\% noise, the PV method yields BER~$\approx 0.7\%$ while the radial integration method achieves BER~$=0$, and at 5\% noise the gap widens to $\approx 18\%$ (PV) versus $\approx 8.9\%$ (Radial). However at 16 modes and 10\% noise, it has opposite result.

Fig.~4(a2,~b2) reveals that a narrower amplitude range increases BER by decreasing the separation between decision boundaries. At 128 modes and 2\% noise, increasing the range from $\pm0.1$ to $\pm0.8$ reduces BER from $\approx 32\%$ to $\approx 0\%$ in PV and from $\approx 22\%$ to $\approx 0\%$ in radial integration. In the result of (a1,~b1) 2\% noise condition, radial integration results have low BER for almost all conditions, which is consistent with the result of (a2,~b2). 
Fig.~4(a3,~b3) examines the effect of modulation depth: as the PAM order increases from 2-PAM to 8-PAM, the amplitude spacing between symbols narrows, leading to sharply higher BER. At 128 modes and 2\% noise, BER rises from $\approx 0.7\%$ (2-PAM) to $\approx 21\%$ (4-PAM) and $\approx 33\%$ (8-PAM) for the PV method. This trade-off between aggregate data-rate and reliability is mitigated in the scenario considering radial distribution, where the broader averaging area provides greater noise tolerance. These results collectively establish the practical operating limits of our method in terms of maximum mode count, tolerable noise level, and achievable modulation order. We adopt 128 ($=2^7$) pair modes as the baseline scenario because it provides an aggregate data-rate of 128 bits/frame at 2-PAM, representing a practical balance between throughput and reliability. The maximum mode count is bounded by the azimuthal sampling resolution: pair mode $p$ requires at least $4p$ pixels along the circumference (Nyquist criterion). At radius $r$ with pixel pitch $\delta$, the circumference provides $2\pi r/\delta$ samples, yielding an upper bound of $p_{\max}\leq\pi r/(2\delta)$. In our simulation ($r=500\ \mu$m, $\delta=1\ \mu$m), this gives $p_{\max}\approx 785$, confirming that the 128 mode scenario operates well within the detector resolution limit. The scalable mode count therefore ranges up to approximately 785 for the given detector geometry, with BER performance degrading gradually as shown in Fig.~4.
Beyond the quantitative analysis, the practical merit of our method lies in its minimal hardware and computational requirements. The entire demultiplexing procedure requires only a CCD sensor without additional optical elements such as spatial light modulators (SLMs), gratings, or mode sorters, so the full incident photon flux reaches the detector without diffractive or absorptive intensity losses. The analytical process consists solely of a subtraction followed by a single FT, eliminating the need for iterative training or numerical optimization and enabling frame rate processing suitable for real-time communication. Noise suppression arises naturally at each stage of this procedure: the subtraction removes common mode noise, the radial integration averages spatially uncorrelated noise over a large number of pixels with a statistical advantage proportional to $\sqrt{N_{\text{number of pixels}}}$, and the FT inherently rejects non harmonic components. Moreover, since mode separation operates in the frequency domain, adding more channels does not introduce inter channel interference, allowing the method to scale linearly with available azimuthal resolution.
We note that the present analysis is conducted entirely through numerical simulation with idealized beam models and additive white noise.
\begin{figure}[!tb]
    \centering
    \includegraphics[width=\linewidth]{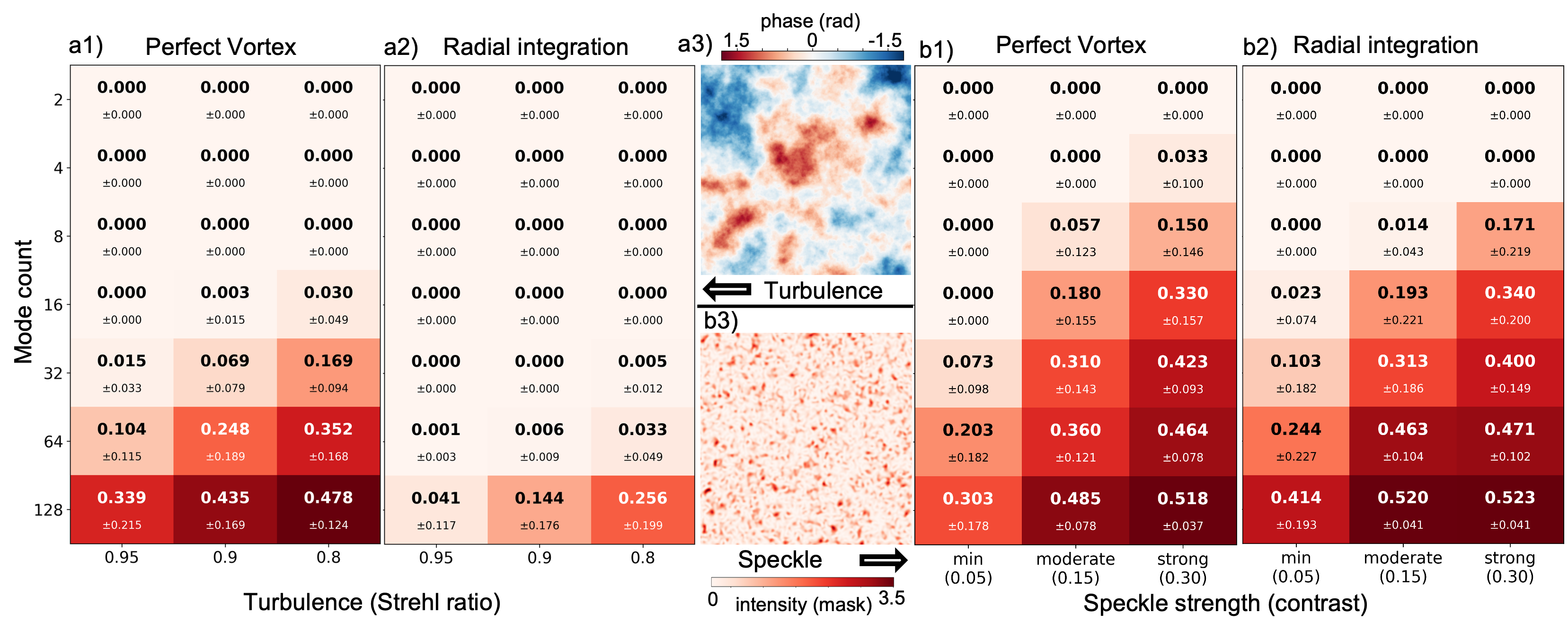}
    \caption{Bit Error Rate (BER) analysis under weak turbulence (a1--a3) and speckle (b1--b3). a1,~b1) BER for the PV scenario versus pair mode count and perturbation strength (turbulence Strehl ratio 0.95/0.9/0.8; speckle contrast 0.05/0.15/0.30). a2,~b2) the same for the radial distribution considered method. a3,~b3) an example turbulence phase screen (SR $=0.8$) and speckle mask ( grain $30 \mu$m, contrast $0.30$), each shown at the strongest condition. Each BER result is averaged over 20 repeated calculations with identical pre-generated random bit streams shared across all conditions. Standard deviations of 20 trials are also calculated.}
    \label{fig:ber_calc2}
\end{figure}

Additionally, we present the effect of weak turbulence and speckle noise on our communication scenario by BER analysis in Fig. 5. We model turbulence as a single Kolmogorov/von Kármán phase screen \cite{Lane1992} distributed over the entire beam aperture, following the Strehl-ratio (SR) formulation \cite{Reis2025}, with its wavefront variance fixed to $\sigma_\phi^2 = 1/(\text{SR}) - 1$ so that the turbulence strength is set solely by the Strehl ratio. Since the beam is compact in our simulation, turbulence would otherwise act on it as a predominantly tip/tilt aberration. Therefore we set the phase-screen correlation length smaller than the beam width, so that the wavefront perturbation is structured (higher-order) and its influence on the demultiplexing is clearly revealed. As the Strehl ratio decreases, the BER increases monotonically while the low-order modes remain robust, and the radial-integration method is more tolerant than the PV circular scan because the azimuthal integration averages out the turbulence-induced intensity fluctuation. We further consider speckle noise from a scattering channel, which multiplies the observed intensity by a granular pattern, $I_{\text{obs}} = I_{\text{clean}}\cdot(\text{mask})$, where the mask is the intensity of a correlated complex-Gaussian field normalized to unit mean \cite{Goodman1976, Duncan2008}. The speckle is controlled by two parameters, the grain size (correlation length) and the contrast $C$ ($=\text{std}/\text{mean}$, $0\le C\le 1$). Since a grain of size $g$ on a ring of radius $R$ produces azimuthal harmonics up to $m\sim 2\pi R/g$, a larger grain contaminates the low-order modes first, whereas a finer grain reaches higher-order modes. For a fine speckle (grain $=30\ \mu$m), the BER increases smoothly with both mode count and contrast. Throughout, reconstruction at the detection plane renders the phase aberration as an intensity perturbation. These results demonstrate the limitations of our method under weak turbulence and speckle noise, and provide a basis for future work. A deep-learning-based intensity-profile reconstruction could act synergistically with our method to improve OAM communication under turbulence and various noise conditions.

\section{Conclusion}
We have presented a demultiplexing method for OAM light that operates entirely within a single intensity profile, eliminating the need for additional optical elements such as gratings, SLMs, or mode sorters. The core principle is orthogonality restoration: by subtracting half of the maximum at each radius or mean intensity from the superposed pair mode profile, the $\cos^2(p\phi)$ basis is transformed into $\cos(2p\phi)$, restoring mutual orthogonality among all pair channels and enabling direct mode separation via a single FT operation.
Numerical simulations with 128 multiplexed pair channels confirmed the method's validity under both PV and considering radial distribution. The PV scenario achieved successful demultiplexing under both ideal condition and white noise 2\% induced condition. The approach of considering radial distribution derived by same principle consistently achieves stable SNR. This arises from the larger spatial coverage of the detection area. Parametric analysis established quantitative operating boundaries in terms of noise level, modulation order, PAM order, turbulence, and speckle noise. Compared to computational approaches, our method relies on a deterministic analytical procedure (followed by FT) rather than trained models, offering transparency and reproducibility.

The primary limitation lies in the pair-mode constraint ($|\ell\rangle + |-\ell\rangle$). Since each channel combines two OAM states of opposite sign, the effective channel count is $p$ rather than $2\ell+1$, reducing number of channels by approximately a factor of two compared to methods that address individual $\ell$ states. Notably, an arbitrary phase error leaves the FT magnitude unchanged, preserving exact amplitude recovery. While the present simulations assume a CCD based detection scheme, the inherent frame rate limitation of imaging sensors imposes a practical throughput ceiling for high-rate communication performance. In the PV scenario, the azimuthal sampling can equivalently be realized by a circular array of discrete photodetectors at the detection radius. This replaces the CCD with a small number of fast photodetectors, removing the frame rate problem while keeping the same restoration and FT processing. Another limitation is that the beam profile drifts with temperature and turbulence, a calibration method has to be considered for practical implementation.
These features establish the proposed scheme as a hardware minimization and analytically transparent framework for OAM based optical communication. This work provides a practical simplification of the demultiplexing method for scalable OAM based communication systems.

\section*{}

\noindent\textbf{Funding.} This work was supported by the Ministry of Science and ICT (MSIT), Korea, through Information Technology Research Center (ITRC) Program (IITP-2026-RS-2023-00259061); Institute for Information and Communications Technology Planning and Evaluation (IITP).

\noindent\textbf{Disclosures.} The authors declare no conflicts of interest.

\noindent\textbf{Data availability.} Data underlying the results presented in this paper are not publicly available at this time but may be obtained from the authors upon reasonable request.

\end{document}